\documentclass[letterpaper,5p,times,preprint]{elsarticle}
\usepackage[english]{babel} \usepackage[T1]{fontenc} \usepackage[latin1]{inputenc}
\usepackage{graphicx} \usepackage{float}  \usepackage{booktabs} \usepackage{array} \usepackage{amsmath} \usepackage{amsopn} \usepackage{amssymb}  \usepackage[table,dvipsnames]{xcolor}
\usepackage{times}
\usepackage{bm}
\usepackage{dirtytalk}
\usepackage{mhchem}
\usepackage{subcaption}
\usepackage[ruled,vlined]{algorithm2e}
\usepackage{makecell}
\usepackage{float}
\usepackage{placeins}
\usepackage{textcomp} 
\usepackage{geometry} \usepackage{float} \usepackage{gensymb}
\usepackage[breaklinks=true, pdfpagemode=UseOutlines, colorlinks=true, linkcolor=black, filecolor=black, urlcolor=black, citecolor=black, plainpages=false, hypertexnames=false, pdfpagelabels=true, hyperindex=true]{hyperref}
\definecolor{HeaderLine}{RGB}{0, 131, 162}

\usepackage{lineno,hyperref}
\modulolinenumbers[5]

    \usepackage{mathtools}
    \biboptions{numbers,sort&compress}

\begin{document}

    
    \begin{frontmatter}


    \title{Analytical Gradient-Based Optimization of CALPHAD Model Parameters}
    
    \author{Courtney Kunselman$^{a}$}
    \author{Brandon Bocklund$^b$}
    \author{Richard Otis$^c$}
    \author{Raymundo Arr\'{o}yave$^{a,d,e}$}
    \address{$^a$Department of Materials Science and Engineering, Texas A\&M University, College Station, TX 77843}
    \address{$^b$Materials Science Division, Lawrence Livermore National Laboratory, Livermore, CA 94550}
    \address{$^c$Proteus Space Inc., Los Angeles, CA, 90021, USA}
    \address{$^d$Department of Mechanical Engineering, Texas A\&M University, College Station, TX 77843}
    \address{$^e$Department of Industrial and Systems Engineering, Texas A\&M University, College Station, TX 77843}
    
    \cortext[mycorrespondingauthor]{Corresponding author email:cjkunselman18@tamu.edu} 
    \begin{abstract} 
    The calibration of CALPHAD (CALculation of PHAse Diagrams) models involves the solution of a very challenging high-dimensional multiobjective optimization problem. Traditional approaches to parameter fitting predominantly rely on gradient-free methods, which while robust, are computationally inefficient and often scale poorly with model complexity. In this work, we introduce and demonstrate a generalizable framework for analytic gradient-based optimization of the parameters of the CALPHAD model enabled by the recently formalized Jansson derivative technique. This method allows for efficient evaluation of gradients of thermodynamic properties at equilibrium with respect to model parameters, even in the presence of arbitrarily complex internal degrees of freedom. Leveraging these semi-analytic gradients, we employ the conjugate gradient (CG) method to optimize thermodynamic model parameters for four binary alloy systems: Cu-Mg, Fe-Ni, Cr-Ni, and Cr-Fe. Across all systems, CG achieves comparable or superior optimality relative to Bayesian ensemble Markov Chain Monte Carlo (MCMC) with improvements in computational efficiency ranging from one to three orders of magnitude. Our results establish a new paradigm for CALPHAD assessments in which high fidelity data-rich model calibration becomes tractable using deterministic gradient-informed algorithms.
    \end{abstract}


    \end{frontmatter}
    



\section{Introduction}
Emerging challenges in materials design increasingly demand accurate and efficient exploration of phase stability in complex, multicomponent systems. The CALPHAD (CALculation of PHAse Diagrams) method addresses this need by constructing thermodynamic models based on physically informed surrogate descriptions of the Gibbs energy for each phase. These models enable reliable predictions of phase equilibria and transformations across a wide range of compositions and temperatures. The Gibbs energy models -- the key to the CALPHAD approach -- consist of both fundamental thermodynamic physical descriptions and expert-specified parametric forms informed by empirical and computational data. Thus, the overall utility of these models heavily relies on the modeler's ability to choose both an appropriate parametric form and corresponding parameter values which agree well with available data and reasonably interpolate in regions of conditions space with no available benchmark.

Various model forms have been rigorously developed to capture contributions to the Gibbs energy of a phase based on its chemical and physical properties. Common examples applicable to many alloy systems include modified power series expansions for temperature dependence, polynomial basis expansions aimed at reducing correlations between coefficients for composition dependence, and compound energy formalism (CEF) sublattice models for ordering. For these widely applied parametric forms, model selection becomes an art of balancing model simplicity against descriptive power in the context of available data. That is, power and polynomial expansions must be truncated to avoid unnecessary complexity and overfitting while still maintaining extrapolative value, and available data limit the parameters which can be fit (e.g. heat capacity data cannot be used to fit the constant term of a temperature power series expansion). Similarly, CEF models with multiple sublattices can describe complex physical phenomena, but they are more computationally expensive to minimize due to the introduction of additional degrees of freedom and require the fitting of compound energies for which experimental data can be limited or unattainable. Historically, the model selection process has mostly relied on expert knowledge of matching model forms with material properties and/or brute-force trial and error performed on a subset of potential models, coupled with a parameter-fitting step \cite{lukas2007computational}. However, recent developments in statistical tools \cite{bocklund2019espei,honarmandi2019bayesian} are making the model selection process more approachable for the non-expert.

Once the Gibbs energy models for each phase are determined, the degrees of freedom of these models must be calibrated against available data. Empirical measurements of equilibrium activity, thermodynamic properties related to derivatives of the Gibbs energy, and phase boundaries provide touchstones on the convex hull of the energy landscape while computational data establish reference points in experimentally-prohibited regions of configuration space. Parameter fitting thus becomes a multiobjective optimization problem where the objective function is often constructed through a maximum likelihood approach, in which model prediction errors are assumed to be normally distributed at mean zero with observation-specific standard deviations and weights.

Because the objective function is often defined as the output of many energy minimization calculations and the dimension of parameter spaces can get large, accurate, efficient evaluations of the gradients of predicted model errors with respect to parameters have proved difficult to obtain. Consequently, the most popular approach to parameter fitting, weighted nonlinear least squares, implemented in CATCalc \cite{shobu2009catcalc}, MTDATA \cite{davies2002mtdata}, Pandat \cite{cao2009pandat}, Thermo-Calc \cite{jansson1984evaluation} and OpenCalphad \cite{sundman2015opencalphad}, employs gradient-free techniques to iteratively solve the normal equations. Due to their inherent gradient-free design, black-box approaches are also becoming more popular in commercial and open source thermodynamic assessment modules, as evidenced by the use of NOMAD in FactSage's Calphad Optimzer \cite{reisuser} and ESPEI's use of Bayesian ensemble Markov Chain Monte Carlo (MCMC) \cite{bocklund2019espei}. 

Although gradient-free methods provide the flexibility to sidestep the difficulty of robustly differentiating the outputs of equilibrium calculations, gradient-informed optimization schemes could greatly reduce the computational cost of parameter optimization. To this end, Zhang et al. recently demonstrated the use of gradient descent to optimize the Ag-Pd and La-C systems \cite{zhang2025new}. However, their approach appears to have involved specially designed analytic loss functions (i.e. the loss functions did not depend on the outputs of equilibrium calculations), which can only be constructed for systems in which all energy models can be written as functions of composition. In a recent publication \cite{kunselman2024analytically}, the current authors rigorously proved the Jansson derivative technique, an adjoint-like method for computing semianalytic derivatives with respect to external conditions at equilibrium that does not require additional function evaluations or matrix inversions, and speculated that this method could be used to calculate gradients with respect to model parameters at equilibrium for models of arbitrary complexity. 

In this work, we demonstrate the use of Jansson derivatives to enable gradient-informed parameter optimization of CALPHAD models via the conjugate gradient (CG) algorithm. We apply this approach to several binary systems, including Cu-Mg, Fe-Ni, Cr-Ni, and Cr-Fe. The performance of CG is evaluated in terms of efficiency and accuracy and benchmarked against the current standard employed by ESPEI: Bayesian ensemble Markov Chain Monte Carlo (MCMC). Although CG provides a promising foundation for gradient-based optimization in CALPHAD assessments, we recognize it as only a first step. Accordingly, we discuss future directions, emphasizing the potential of stochastic and global optimization strategies to further enhance model calibration.

\section{Theory}
\subsection{Parameter Optimization by Maximum Likelihood Estimation}

A common approach to fitting parametric models involves building a joint probability distribution over parameter space in the context of observed data. In Maximum Likelihood Estimation (MLE), this joint probability distribution describes the likelihood of observing a data sample $(X_1, X_2, \hdots, X_n)$ given a parameter vector $\boldsymbol{\theta}$ and is aptly referred to as the likelihood function. When the sample random variables are assumed to be independent and identically distributed, the likelihood function takes the form 

\begin{equation}
L(\boldsymbol{\theta}) = f(X_1, X_2, \hdots, X_n | \boldsymbol{\theta})=\prod_{i=1}^{n}f_i(X_i | \boldsymbol{\theta})
\end{equation}

\noindent where $f$ is the joint conditional probability density function given $\boldsymbol{\theta}$ of the drawing sample $(X_1, X_2, \hdots, X_n)$ and $f_i$ is the univariate conditional function for observation $X_i$. The MLE estimator, or the most optimal set of parameters, is then found by maximizing $L(\boldsymbol{\theta})$. In practice, the log-likelihood is often maximized in order to avoid floating point precision issues from the multiplication of many small numbers. In addition, in the case of gradient-based optimization, differentiating sums is much simpler than differentiating large products. Thus, the objective function to maximize becomes

\begin{equation} \label{log_likelihood}
l(\boldsymbol{\theta})=\ln{L(\boldsymbol{\theta})}=\sum_{i=1}^n\ln{f_i(X_i | \boldsymbol{\theta})}.
\end{equation}

In this work, we define all $f_i$ as the conditional probability distribution given by the parameter vector $\boldsymbol{\theta}$ of observing a residual (or error) $X_i$. For some data types, $X_i$ is simply the difference between the measured and predicted values while other data types employ alternate error definitions. We also assume that these residuals are normally distributed, and because we want zero error to be most likely, we center these distributions on zero. Standard deviations $\sigma_i$ are data-specific and allow the incorporation of multiple data types by capturing experimental uncertainty and appropriately scaling each contribution to the likelihood function. Each data point can also be assigned a unique weight $w_i$ which divides $\sigma_i$ and provides more user control over the relative importance of each data point. Thus, $\forall i \in\{1,2,\hdots,n\}$

\begin{equation}
    f_i(X_i|\boldsymbol{\theta})=\frac{w_i}{\sigma_i\sqrt{2\pi}}\exp\left({-\frac{1}{2}\left(\frac{w_iX_i}{\sigma_i}\right)^2}\right),
\end{equation}

\noindent which further implies

\begin{equation} \label{contribution}
    \ln{f_i(X_i|\boldsymbol{\theta})} = \ln{\frac{w_i}{\sigma_i\sqrt{2\pi}}}  -\frac{1}{2}\left(\frac{w_iX_i}{\sigma_i}\right)^2.
\end{equation}

Since we are performing gradient-based optimization, we also need an expression for the derivative of $l(\boldsymbol{\theta})$ with respect to $\boldsymbol{\theta}$. Plugging Eq. \ref{contribution} into Eq. \ref{log_likelihood} and carrying out the prescribed differentiation gives

\begin{equation} \label{log_likelihood_gradient}
    \frac{\partial l(\boldsymbol{\theta})}{\partial \boldsymbol{\theta}}=\sum_{i=1}^n-\frac{w_i^2X_i}{\sigma_i^2}\frac{\partial X_i}{\partial \boldsymbol{\theta}}.
\end{equation}

\noindent Historically, the difficulty in evaluating Eq. \ref{log_likelihood_gradient} lies in the calculation of $\frac{\partial X_i}{\partial \boldsymbol{\theta}}$. For example, let $X_i$ be the residual between an empirically observed enthalpy value for a closed system at a given temperature $T$ and composition $\boldsymbol{x}$ for phase $\alpha$ and the model's predicted enthalpy under such conditions. Assume that phase $\alpha$ is modeled using the Compound Energy Formalism (CEF) \cite{hillert2001compound} and that it has multiple sublattices, making the enthalpy per mole formula unit of phase $\alpha$ a function of temperature and site fraction $\boldsymbol{y}_s$ for each sublattice $s$. Because site fractions are degrees of freedom of the Gibbs energy minimization calculation, they are functions of $\boldsymbol{\theta}$, but the minimization procedure often does not produce an explicit closed-form expression of site fractions as functions of $\boldsymbol{\theta}$. Thus, to fully capture how enthalpy changes with changing $\boldsymbol{\theta}$, we need a way to quantify how site fractions change with $\boldsymbol{\theta}$.

This is where the Jansson derivative technique comes in. As mentioned in the introduction, Jansson derivatives leverage an adjoint-like approach to compute derivatives of the degrees of freedom of Gibbs energy minimization calculations with respect to external conditions of the equilibrium calculation. As shown in \cite{kunselman2024analytically}, Jansson derivatives are queries of the analytic derivative at the point in the conditions space at which the equilibrium calculation is carried out, and require no additional equilibrium calculations (or even matrix inversions), making them both highly accurate and extremely computationally efficient compared to numerical approaches. When model parameters are treated as potential conditions (similar to temperature or pressure), the Jansson derivative method allows efficient calculation of analytic gradients of equilibrium thermodynamic properties with respect to $\boldsymbol{\theta}$, which in turn facilitates the calculation of Eq. \ref{log_likelihood_gradient} for use in gradient-informed optimization approaches.

\subsection{Data Types, Likelihood Contributions, and Gradients}

ESPEI can optimize CALPHAD model parameters using four main data types: empirical equilibrium thermochemical data, computational non-equilibrium thermochemical data, empirical activity data, and empirical zero phase fraction (phase equilibria) data. The contributions to the overall likelihood function and the gradients of each of these data types are described below.

\subsubsection*{Equilibrium Thermochemical Data}

Equilibrium thermochemical data refer to single-phase measurements of Gibbs energy or its derivatives in which internal degrees of freedom of the phase are at their equilibrium values given the supplied external conditions, but the equilibrium configurations themselves are not specified. For this data type, the residual $X^{eq}_i$ is simply the difference between the calculated property $Z^{eq}_{i,calc}$ produced by an equilibrium calculation performed on the present model and the observed property $Z^{eq}_{i,obs}$:

\begin{equation}
    X^{eq}_i = Z^{eq}_{i,calc} - Z^{eq}_{i,obs}.
\end{equation}

\noindent This implies that the gradient of the residual is equivalent to the gradient of the calculated property, and we compute it using the Jansson derivative method at the conclusion of the Gibbs energy minimization:

\begin{equation} \label{eq_grad}
    \frac{\partial X^{eq}_i}{\partial\boldsymbol{\theta}}=\frac{\partial Z^{eq}_{i,calc}}{\partial\boldsymbol{\theta}}.
\end{equation}

Presently, PyCalphad can only calculate first-order Jansson derivatives. Consequently, at first glance it would appear that Eq. \ref{eq_grad} can only be calculated for residuals of the Gibbs energy and not for any of its derivatives. However, as shown in \cite{kunselman2024analytically} for closed systems, equilibrium properties derived from first derivatives of the Gibbs energy with respect to temperature and pressure can be calculated without the need for Jansson derivatives (i.e. contributions to the total derivative from the degrees of freedom of the equilibrium calculation cancel out). Thus, our framework can presently incorporate equilibrium measurements of entropy and enthalpy, but isobaric heat capacity inputs are on hold until second-order derivative capability is implemented. 

\subsubsection*{Non-equilibrium Thermochemical Data}

As opposed to the previous data type, non-equilibrium thermochemical data encompass single-phase values of the Gibbs energy and its derivatives where the internal degrees of freedom of the phase are fixed. Similarly to the previous data type, the residual $X^{neq}_i$ is the difference between the calculated property $Z^{neq}_{i,calc}$ and the observed property $Z^{neq}_{i,obs}$:

\begin{equation}
    X^{neq}_i = Z^{neq}_{i,calc} - Z^{neq}_{i,obs}.
\end{equation}

\noindent However, because the internal degrees of freedom of the phase are specified (and, accordingly, there is no requirement for the system to be in an equilibrium state), $Z^{neq}_{i,calc}$ can be computed by performing any necessary partial derivatives on the Gibbs energy model of the phase followed by directly plugging in the given conditions and internal degrees of freedom. The residual gradient is once again equivalent to the calculated property gradient:

\begin{equation} \label{neq_grad}
    \frac{\partial X^{neq}_i}{\partial\boldsymbol{\theta}}=\frac{\partial Z^{neq}_{i,calc}}{\partial\boldsymbol{\theta}},
\end{equation}

\noindent but because the calculation of $Z^{neq}_{i,calc}$ does not require an equilibrium calculation, its gradient is determined simply by computing the partial derivatives of the model of $Z$ with respect to the parameters. This means that our gradient-based framework can incorporate non-equilibrium thermochemical data that require second-order derivatives of the Gibbs energy, such as isobaric heat capacity.

\subsubsection*{Activity Data}

Because activity data are a non-linear transformation of one of the derivatives of the Gibbs energy, we construct its contribution to the overall likelihood function differently than for other equilibrium thermochemical data.
That is, we are assuming that the residuals of equilibrium chemical potentials are normally distributed, not the residuals of activities. Therefore, to begin building the residual, we first need to convert the observed activities $a_{i,obs}$ into chemical potentials $\mu_{i,obs}$ using the relation:

\begin{equation} \label{conversion}
    \mu_{i,obs}=\mu^0_i + RT\ln{a_{i,obs}}
\end{equation}

\noindent where $\mu^0_i$ is the chemical potential in the reference state for the activity data point $i$, $R$ is the ideal gas constant and $T$ is the temperature. The residual $X^{acr}_{i}$ is then the difference between the predicted chemical potential of the model $\mu_{i,calc}$ and $\mu_{i,obs}$:

\begin{equation} \label{acr_residual}
    X^{acr}_{i} = \mu_{i,calc} - \mu_{i,obs}.
\end{equation}

Before presenting the gradient of $X^{acr}_i$, we must point out an important note about Eq. \ref{acr_residual}. That is, the activity data set only provides the phase and external conditions for the reference state chemical potential; $\mu^0_i$ itself is determined through a single-phase equilibrium calculation using the current model and these given conditions. On the other hand, $\mu_{i,calc}$ is determined by an equilibrium calculation with conditions identical to those of the activity measurement, and all phases are considered. With this in mind, Eqs. \ref{conversion} and  \ref{acr_residual} tell us that the gradient of $X^{acr}_i$ with respect to $\boldsymbol{\theta}$ is

\begin{equation}
    \frac{\partial X^{acr}_i}{\partial\boldsymbol{\theta}}=\frac{\partial\mu_{i,calc}}{\partial\boldsymbol{\theta}} - \frac{\partial\mu^0_{i}}{\partial\boldsymbol{\theta}}
\end{equation}

\noindent where the derivatives of $\mu_{i,calc}$ and $\mu^0_{i}$ are calculated using the Jansson derivative method on the end products of their separate, respective equilibrium calculations.

\subsubsection*{Zero Phase Fraction (ZPF) Data}

The fourth data type that ESPEI can incorporate into the overall likelihood function is ZPF or phase equilibria data. ESPEI can utilize phase boundary data with an arbitrary number of components and phases (always obeying Gibbs phase rule, of course) and even define residuals for single phase equilibria data. Furthermore, phases of interest do not have to be stable under the given conditions in order to have a defined residual with the current selection of model parameters, which is not possible when the residual is defined as the difference between calculated versus observed values (as with the previous three data types). This flexibility is afforded by defining the residual $X^{zpf}_{i,\alpha}$ as the driving force for phase $\alpha$ to be stable at its observed temperature and vertex composition $\boldsymbol{x}^\alpha$:

\begin{equation} \label{residual_driving_force}
    X^{zpf}_{i,\alpha} = \sum_A\bar{\mu}_Ax^\alpha_A-G^\alpha
\end{equation}

\noindent where $\bar{\mu}_A$ is the chemical potential of component $A$ corresponding to the target hyperplane, $x^\alpha_A$ is the composition of component $A$ at the observed $\alpha$ phase vertex and $G^\alpha$ is the minimum Gibbs energy of the $\alpha$ phase conditioned on $\boldsymbol{x}^\alpha$. Notably, this metric, named the residual driving force, is employed with slight differences in a few previous studies \cite{bocklund2019espei,otis2017high,otis2021sensitivity}, but the definition provided in Eq. \ref{residual_driving_force} is aligned with the description given in \cite{bocklund2021computational,kunselman2024analytically}. 

In this work, the target hyperplane chemical potentials are determined by averaging the chemical potentials from multiphase equilibrium calculations conducted at all phase vertex compositions provided in the data. If the data identify the phase as stable but its vertex composition is unreported, that phase does not contribute to the target hyperplane. However, we still calculate a residual for these phases. This is accomplished by sampling the corresponding Gibbs energy model at 50 different points in configuration space and choosing the internal degree-of-freedom configuration (and related vertex composition) which maximizes the residual driving force. Note that Eq. \ref{residual_driving_force} is written so that $X^{zpf}_{i,\alpha}$ is negative when $G^\alpha$ is above the target hyperplane, so maximizing the residual driving force is equivalent to choosing the configuration which is most stable relative to the target hyperplane.

A more elegant and consistent method for determining the vertex composition which minimizes the driving force of a phase given a target hyperplane would be to conduct a single-phase equilibrium calculation conditioned on the chemical potentials of said hyperplane (this would require any composition conditions and the assumption that the system is closed to be relaxed). The energy minimization procedure would then provide the composition which maximizes the residual driving force. PyCalphad does not currently support this kind of equilibrium calculation, but we look forward to its future implementation.

Regardless of how the vertex compositions are chosen, the gradient of $X^{zpf}_{i,\alpha}$ with respect to $\boldsymbol{\theta}$ is

\begin{equation} \label{zpf_grad_unsimplified}
    \frac{\partial X^{zpf}_{i,\alpha}}{\partial \boldsymbol{\theta}} = \sum_A\left(\frac{\partial\bar{\mu}_A}{\partial\boldsymbol{\theta}}x^\alpha_A+\bar{\mu}_A\frac{\partial x^\alpha_A}{\partial\boldsymbol{\theta}}\right) - \frac{\partial G^\alpha}{\partial \boldsymbol{\theta}} - \sum_{A,A\neq B}\frac{\partial G^\alpha}{\partial x^\alpha_A}\frac{\partial x^\alpha_A}{\partial\boldsymbol{\theta}}
\end{equation}

\noindent where $B$ is the dependent component. The derivatives of $\bar{\mu}$ and $G^\alpha$ with respect to the model parameters are calculated using the Jansson derivative method at the conclusion of their respective equilibrium calculations (the arithmetic mean is a linear operator, implying that the derivative of the mean is the mean of the derivatives). When $\boldsymbol{x}^\alpha$ is reported in the data, it is treated as a constant and $\frac{\partial x^\alpha_A}{\partial \boldsymbol{\theta}}=\boldsymbol{0}$ for all components $A$. If $\boldsymbol{x}^\alpha$ is determined by sampling the configuration space, $\boldsymbol{x}^\alpha$ becomes a function of $\boldsymbol{\theta}$. However, due to the discrete nature of the sample, we can think of each $x^\alpha_A(\boldsymbol{\theta})$ as a multidimensional step function with 50 possible ``steps." Thus, for all possible $\boldsymbol{\theta}$, $\frac{\partial x^\alpha_A}{\partial \boldsymbol{\theta}}$ is either $\boldsymbol{0}$ or undefined. In practice, we make the assumption that we will never land on a set of parameters which makes $x^\alpha_A$ undefined, and we set $\frac{\partial x^\alpha_A}{\partial \boldsymbol{\theta}}=\boldsymbol{0}$. Thus, for all of the data in our current implementation, Eq. \ref{zpf_grad_unsimplified} simplifies to

\begin{equation}
    \frac{\partial X^{zpf}_{i,\alpha}}{\partial \boldsymbol{\theta}} = \sum_A\frac{\partial\bar{\mu}_A}{\partial\boldsymbol{\theta}}x^\alpha_A - \frac{\partial G^\alpha}{\partial \boldsymbol{\theta}}.
\end{equation}

If the method of using an equilibrium calculation conditioned on target hyperplane chemical potentials is used to determine the vertex composition of phase $\alpha$, then all $x^\alpha_A$ would once again be functions of $\boldsymbol{\theta}$. In this case, the gradients of the vertex compositions with respect to model parameters would not all be the zero vector, but they could be determined by using the Jansson derivative technique on the end products of said equilibrium calculation.

\subsection{Optimization Procedure}

There are a plethora of gradient-based optimization algorithms available, ranging from simple, deterministic local routines such as gradient descent to more complicated, stochastic global approaches such as Adam \cite{kingma2014adam} or NUTS \cite{hoffman2014no}, a recent adaptation of Hamiltonian Monte Carlo (HMC). In this first work employing Jansson-derivative-enabled gradient-based techniques for CALPHAD model parameter optimization, we decided to explore the use of the former using scipy's CG implementation \cite{2020SciPy-NMeth}. CG is a modified version of gradient descent which, in the linear case, determines $N$ orthogonal search directions in a stretched parameter space (where $N$ is the number of parameters) and takes a maximum of one step along each search direction to find the local optima. Nonlinear implementations operate similarly, but they require ``restarts" to the set of search directions as they lose conjugacy. The step size for each iteration is adaptive and is theoretically chosen such that the objective function is minimized along the search direction. In practice, a line search method queries the objective function and its gradient along the specified search direction at various step sizes until it finds a step size which ensures 1) sufficient decrease of the objective and 2) sufficient increase or flattening of the directional derivative. Thus, the efficiency and robustness of nonlinear implementations can heavily rely on the specific line search method and user-specified hyperparameters describing what changes in the objective and its gradient are considered ``sufficient."  

These implementations are flexible enough to work with any continuous, differentiable objective function, but tend to enjoy the most success when the gradient of the objective is approximately linear near local optima. All contributions to the gradient of the likelihood function from non-equilibrium thermochemical data will be linear in the parameters, but contributions from the other three data types can exhibit non-linear behavior in some areas of parameter space due to changes in the set of stable phases, changes in the internal phase degrees of freedom, or the derivatives of vertex composition with respect to parameters being non-zero for ZPF data. We expect nonlinear contributions to the gradient to be sparse in regions near local optima, which gives us confidence that nonlinear CG will not skip over local optima. 

Because of large disparities in the contributions to the objective function due to small changes in each degree of freedom throughout the domain, preconditioning, or stretching of the local quadratic form of the objective to make it more spherical, is essential for CG's efficiency and performance. As recently stated by Andrei in \cite{andrei2020nonlinear}, proper CG preconditioning in the linear context is well understood, but remains an open question for the nonlinear application. Andrei shows that dynamic approximations to the inverse of the Hessian of the objective at each iteration can be a good choice for nonlinear preconditioners, but for expensive objective functions in large-dimensional spaces, numerically computing these approximations can become computationally intractable. As mentioned in the previous paragraph, in this work we are expecting many of the contributions to the likelihood function to be linear. Also, because we are using an off-the-shelf CG implementation, we desire a computationally-inexpensive, static preconditioning strategy. Therefore, we adopt a heuristic Jacobi-like \cite{shewchuk1994introduction} preconditioning approach where we simply scale the coordinate axes to induce the desired stretching of the objective function.

Lastly, it is important to note that our goal is not to claim that deterministic local routines are the cutting edge answer for arbitrarily complex CALPHAD parameter fitting problems, but rather to show that the general Jansson derivative formulation enables us to implement and demonstrate an efficient and robust framework for calculating the likelihood function and its derivatives so that modelers can choose a gradient-based method which best fits their system and data.

\section{Parameter Optimization Method Comparison}
In this section, we compare the performance of CG against that of Bayesian ensemble MCMC in optimizing the parameters for the Cu-Mg, Fe-Ni, Cr-Ni, and Cr-Fe systems. For all four systems, model and starting point selection was accomplished via ESPEI's parameter generation feature given the sublattice models provided in Table \ref{tab:sublattice models}. Following the approach of Bocklund \cite{bocklund2021computational}, if entropy data was unavailable to fit temperature-dependent parameter coefficients, it was estimated using the method proposed by Witusiewicz and Sommer \cite{witusiewicz2000estimation}: $\Delta H \approx 3000\Delta S$. MCMC runs were conducted using ESPEI v0.9.0, while the likelihood function and its gradient were computed for CG using ESPEI pull request \#268 \cite{espei_268}. We note that there is a slight difference between these two versions in how the ZPF residual (Eq. \ref{residual_driving_force}) is calculated when the vertex composition is not given. In the released version, $G^\alpha$ is computed at the sampled configuration which maximizes the residual driving force (there is no subsequent energy minimization calculation to place it on the convex hull). Thus, differences arise only for phases with internal degrees of freedom, and the overall likelihoods tend to differ by less than two percent. For all final comparisons of optimality across the two methods, all likelihoods are calculated using ESPEI pull request \#268.

In its CG implementation, scipy uses the line search method presented in \cite{nocedal1999numerical} based on the strong Wolfe conditions with default values of $c_1=10^{-4}$ and $c_2=0.4$ for the sufficient decrease and curvature conditions, respectively. Unless otherwise indicated, we use the default hyperparameters for the optimization of all four systems. 

\begin{table}[h]
    \centering
    \begin{tabular}{lll}
       System & Phase & Sublattice Model\\
       \hline
       Cu-Mg & Liquid & (Cu, Mg)\textsubscript{1}\\
             & FCC & (Cu, Mg)\textsubscript{1}\\
             & HCP & (Cu, Mg)\textsubscript{1}\\
             & CuMg\textsubscript{2} & (Cu)\textsubscript{1}(Mg)\textsubscript{2}\\
             & Laves C15 & (Cu, Mg)\textsubscript{2}(Cu, Mg)\textsubscript{1}\\
             \hline
      Cr-Fe & Liquid & (Cr, Fe)\textsubscript{1}\\
            & BCC & (Cr, Fe)\textsubscript{1}\\
            & FCC & (Cr, Fe)\textsubscript{1}\\
            & $\sigma$ & (Cr, Fe)\textsubscript{10}(Cr, Fe)\textsubscript{4}(Cr, Fe)\textsubscript{16}\\
            \hline
      Cr-Ni & Liquid & (Cr, Ni)\textsubscript{1}\\
            & BCC & (Cr, Ni)\textsubscript{1}\\
            & FCC & (Cr, Ni)\textsubscript{1}\\
            \hline
      Fe-Ni & Liquid & (Fe, Ni)\textsubscript{1}\\
            & BCC & (Fe, Ni)\textsubscript{1}\\
            & FCC & (Fe, Ni)\textsubscript{1}\\
            & FCC L1\textsubscript{2} & (Fe)\textsubscript{1}(Ni)\textsubscript{3}
            
    \end{tabular}
    \caption{Sublattice models for each phase by system.}
    \label{tab:sublattice models}
\end{table}

\subsection{Data Description and Model Selection}

The Cu-Mg system has been previously assessed using MCMC \cite{bocklund2019espei,bocklund2021computational} and is featured in ESPEI's user tutorial, making it an excellent candidate system for optimization method comparison. This system consists of four solution phases (liquid, FCC, HCP, and a Cu\textsubscript{2}Mg C15 Laves phase) and one stoichiometric compound, CuMg\textsubscript{2}. The empirical single-phase thermochemical data employed for parameter generation are composed of mixing enthalpies for the liquid phase \cite{batalin1987thermodynamic, sommer1983calorimetric} and formation enthalpies for the CuMg\textsubscript{2} phase and the Laves phase Cu:Mg end-member \cite{feufel1995thermodynamic,king1964thermochemical}. Utilized computational thermochemical data from first-principles calculations include enthalpies of mixing for the FCC \cite{gao2014first}, HCP \cite{gao2014first,shin2007thermodynamic}, and Laves phases \cite{bocklund2019espei}, and enthalpies of formation for the endmembers of the Laves and CuMg\textsubscript{2} phases \cite{zhou2007}. Parameter generation produced a model with 22 total degrees of freedom for the binary interaction parameters and end-member compound energies. Details on the exact models and starting points produced through ESPEI's parameter generation feature for all systems can be found in Appendix A in Tables \ref{tab:cu_mg_details}-\ref{tab:cr_fe_details}. Once the model and starting point were determined, phase equilibria data \cite{bagnoud1978binary,jones1931copper,sahmen1908legierungen,urazova1907experimental} were added to the thermochemical data to build the likelihood function for optimization. 

The Fe-Ni system contains four equilibrium phases: liquid, BCC, FCC, and the ordered FCC L1\textsubscript{2} phase. Although L1\textsubscript{2} can be modeled as a solution phase, for simplicity we choose to represent it as a stoichiometric compound. For this system, magnetic contributions to the Gibbs energy need to be taken into account, and they are captured using the Inden-Hillert-Jarl model. The parameters for these magnetic contributions are pulled from \cite{hillert1990reassessment} and remain fixed. Thermochemical data used for parameter generation includes first-principles calculations of the enthalpies of formation for the BCC \cite{chentouf2017,bocklund2021computational}, FCC \cite{bocklund2021computational}, and FCC L1\textsubscript{2} \cite{cacciamani2010,bocklund2019espei} phases and experimental enthalpies of mixing for the FCC \cite{dench1963,kubaschewski1967,steiner1961} and liquid \cite{batalin1974,iguchi1981,predel1970,thiedemann1998} phases. Phase equilibria data \cite{hellawell1957,schurmann1977untersuchungen,cacciamani2010} were also added for the optimization processes. We note that, based on our chosen sublattice model, we modified the vertex composition for all FCC L1\textsubscript{2} data to be the stoichiometric composition. ESPEI's parameter generation module selected a model with nine total degrees of freedom.

Similar to Cu-Mg, the Cr-Ni system was previously assessed using MCMC \cite{otis2021sensitivity,bocklund2021computational} and is featured in the 2021 Materials Genome Workshop demonstration of ESPEI. We model three single-sublattice solution phases for the Cr-Ni system: liquid, BCC, and FCC. For this system, all single-phase thermochemical data for model selection and coefficient fitting are empirical enthalpies of mixing for the BCC \cite{dench1963,watson1995}, FCC \cite{dench1963,watson1995}, and liquid \cite{thiedemann1998,saltykov2002enthalpy} phases. A model with ten degrees of freedom was selected by the parameter generation procedure, and the overall likelihood function was augmented with phase equilibria data from \cite{bechtoldt1961redetermination,collins1988electron,dench1963,jenkins1937some, jette1934x,karmazin1982lattice,svechnikov1962characteristics,taylor1953constitution,zhang2014impurity}.

Lastly, for the Cr-Fe system we consider the liquid, BCC, FCC, and sigma ($\sigma$) phases. Similar to the Fe-Ni system, magnetic contributions to the Gibbs energy are accounted for using the Inden-Hillert-Jarl model, and parameters for such contributions are extracted from \cite{hillert1981some} (excluding the Cr FCC parameters). Single-phase thermochemical data for this system consists of empirical enthalpies of mixing for the BCC \cite{dench1963}, FCC \cite{hultgren1973}, and liquid \cite{batalin1984enthalpies,iguchi1982,thiedemann1998} phases as well as first principles calculations of enthalpies of mixing \cite{kuronen2015} and formation \cite{jiang2004} for the BCC phase and enthalpies of formation \cite{shang2010,jacob2018,pavluu2010ab} for end-members of the sigma phase. We note that, due to a lack of mixing thermochemical data for the sigma phase, ESPEI does not generate any interaction parameters for that phase. Leveraging the parameter generation feature, we selected a model with 26 total degrees of freedom. The overall likelihood function was constructed from the aforementioned thermochemical and additional phase equilibria data given in \cite{dubiel1987miscibility,novy2009atomic,adcock1931alloys,roe1952,zhang2014impurity,belton1970mass,hellawell1957,kundrat1980,schurmann1977untersuchungen}.

\subsection{Optimization Method Comparison}
\begin{figure*}[!htbp]
        \captionsetup[subfigure]{labelformat=empty}
        \centering
        \subfloat[\hspace{.75 in}(a)]{\includegraphics[width=.36\linewidth]{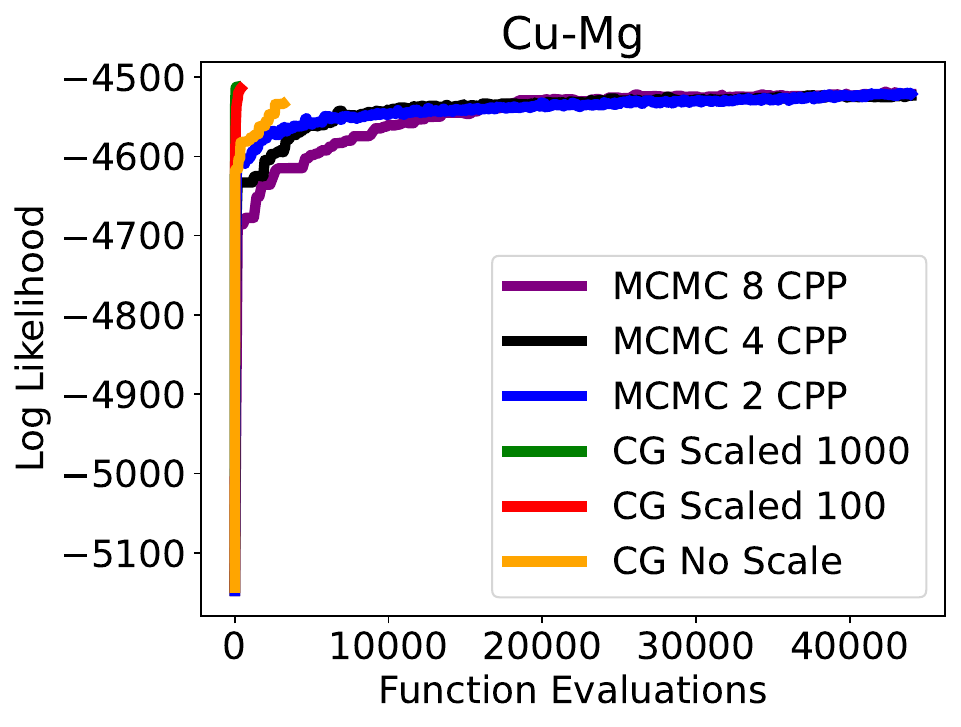}}\hspace{.3 in}
         \subfloat[\hspace{.75 in}(b)]{\includegraphics[width=.36\linewidth]{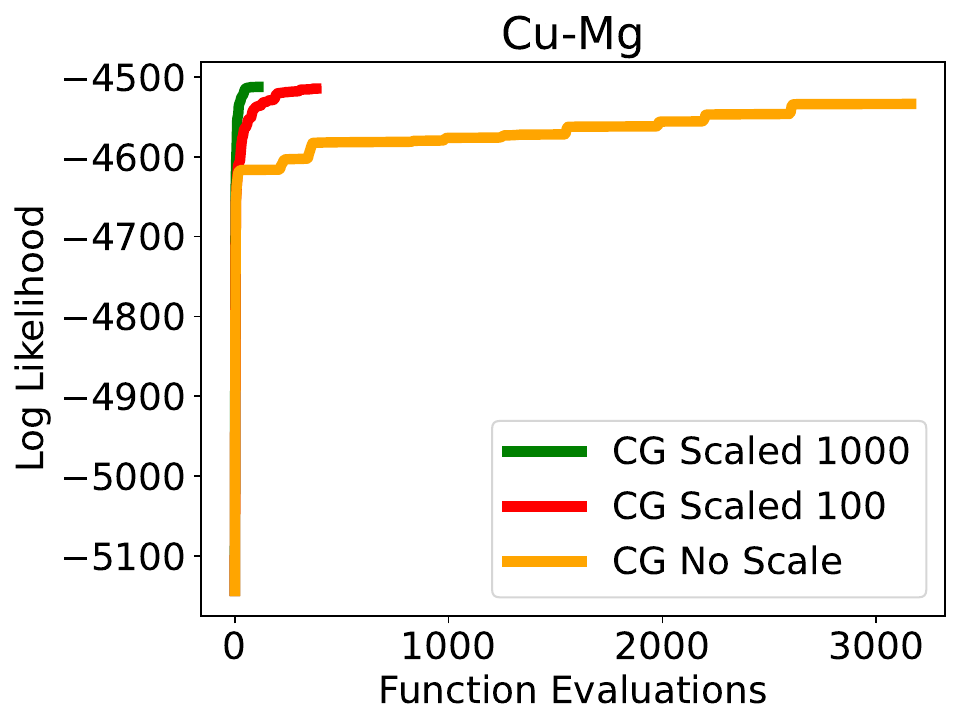}}\\
         \subfloat[\hspace{.75 in}(c)]{\includegraphics[width=.36\linewidth]{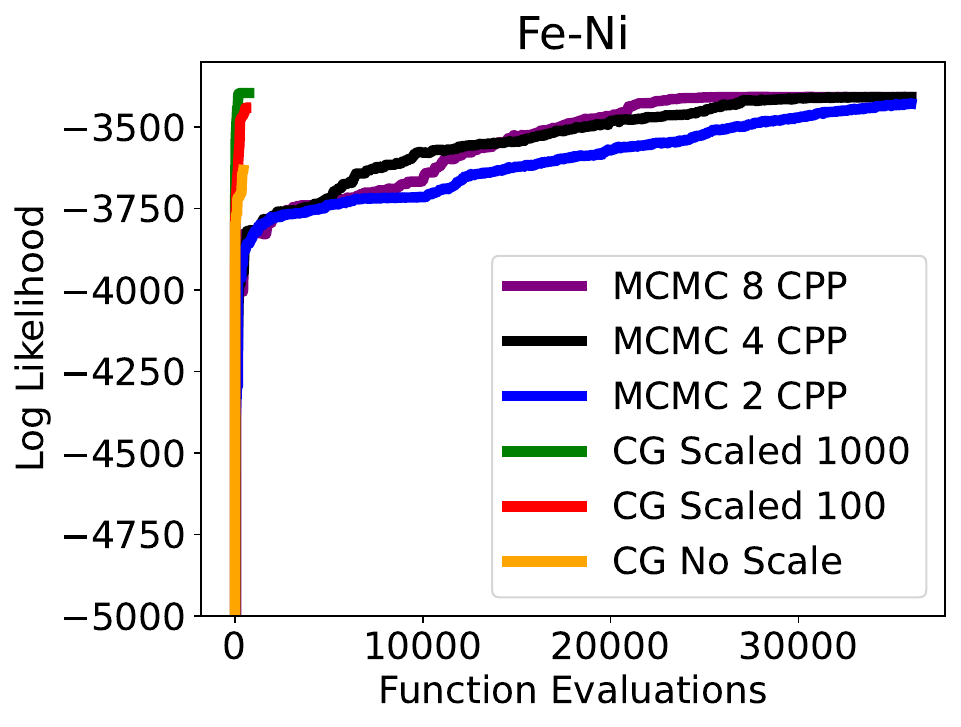}}\hspace{.3 in}
         \subfloat[\hspace{.75 in}(d)]{\includegraphics[width=.36\linewidth]{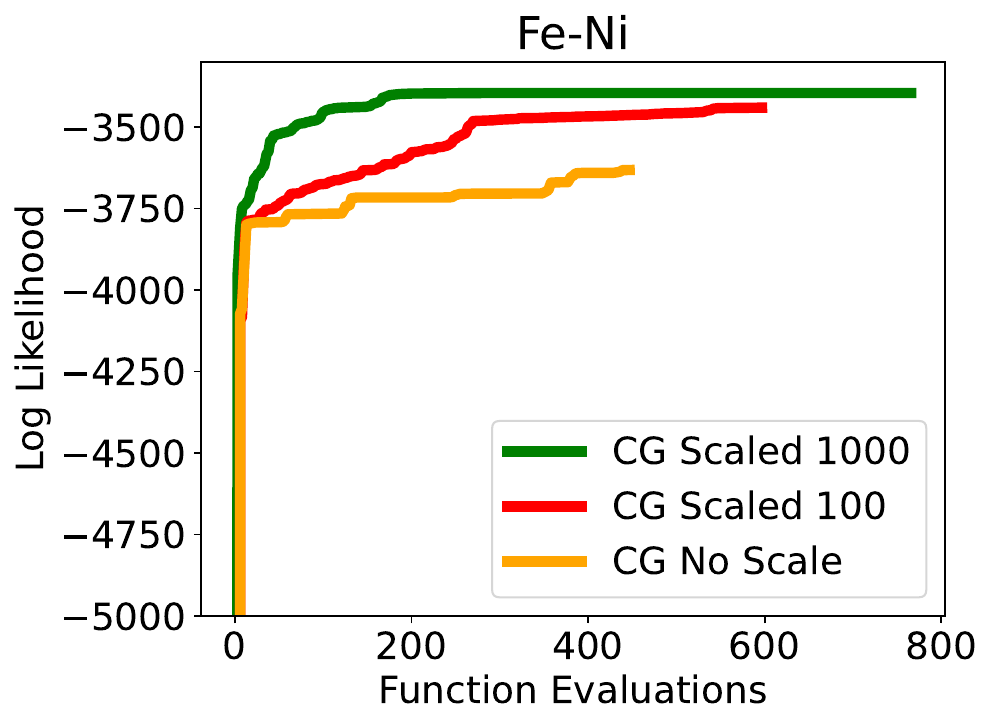}}\\
         \subfloat[\hspace{.75 in}(e)]{\includegraphics[width=.36\linewidth]{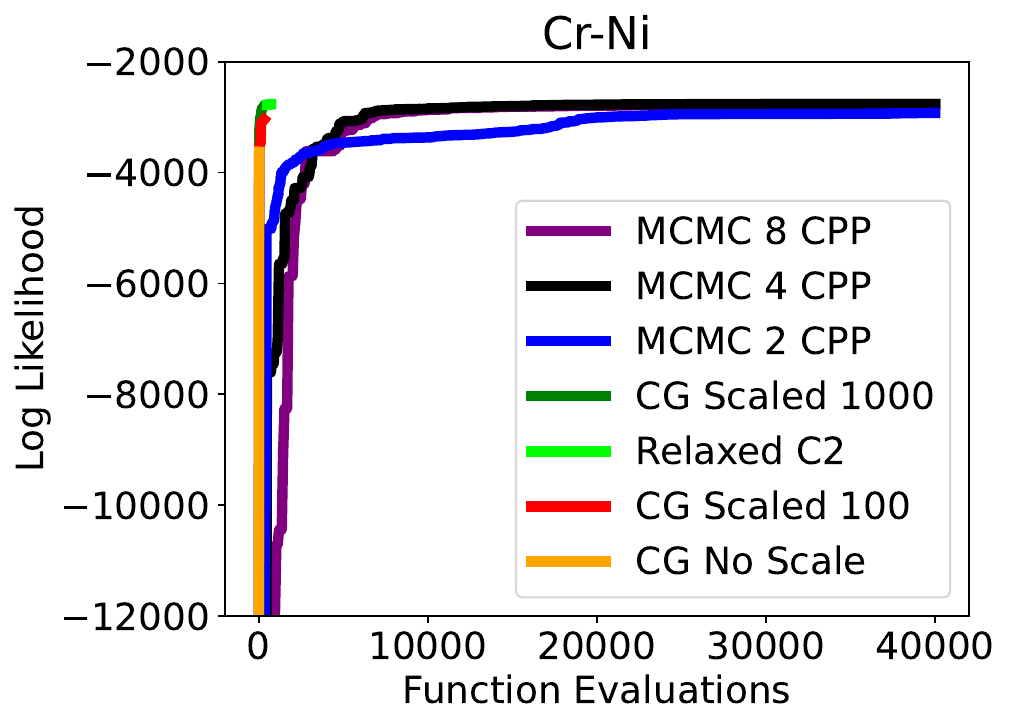}}\hspace{.3 in}
         \subfloat[\hspace{.75 in}(f)]{\includegraphics[width=.36\linewidth]{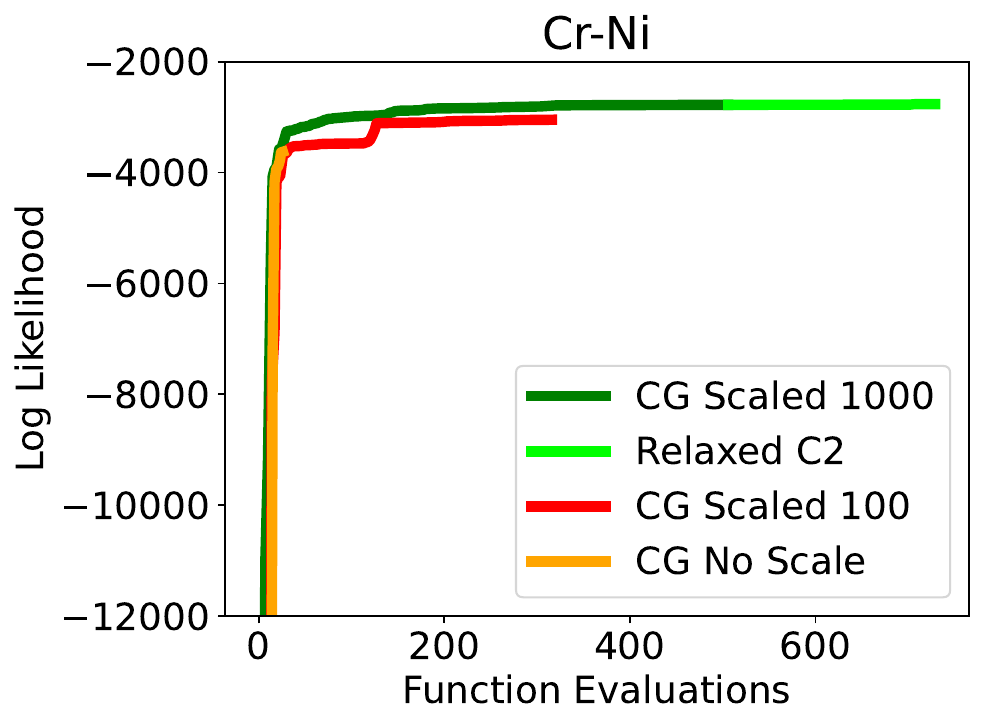}}\\
         \subfloat[\hspace{.75 in}(g)]{\includegraphics[width=.36\linewidth]{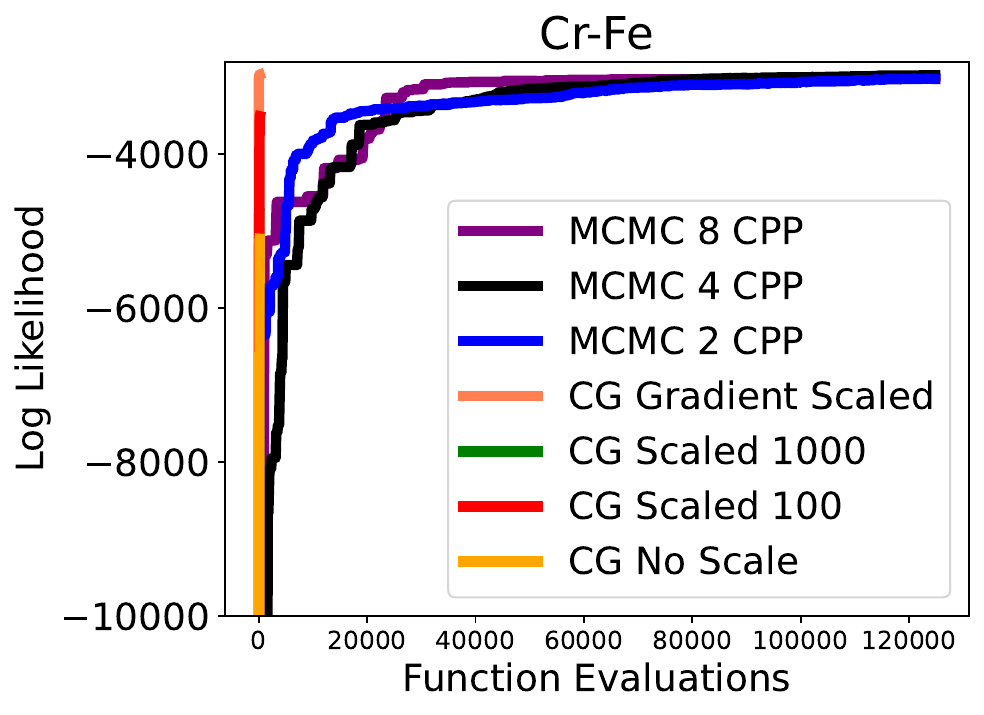}}\hspace{.3 in}
         \subfloat[\hspace{.75 in}(h)]{\includegraphics[width=.36\linewidth]{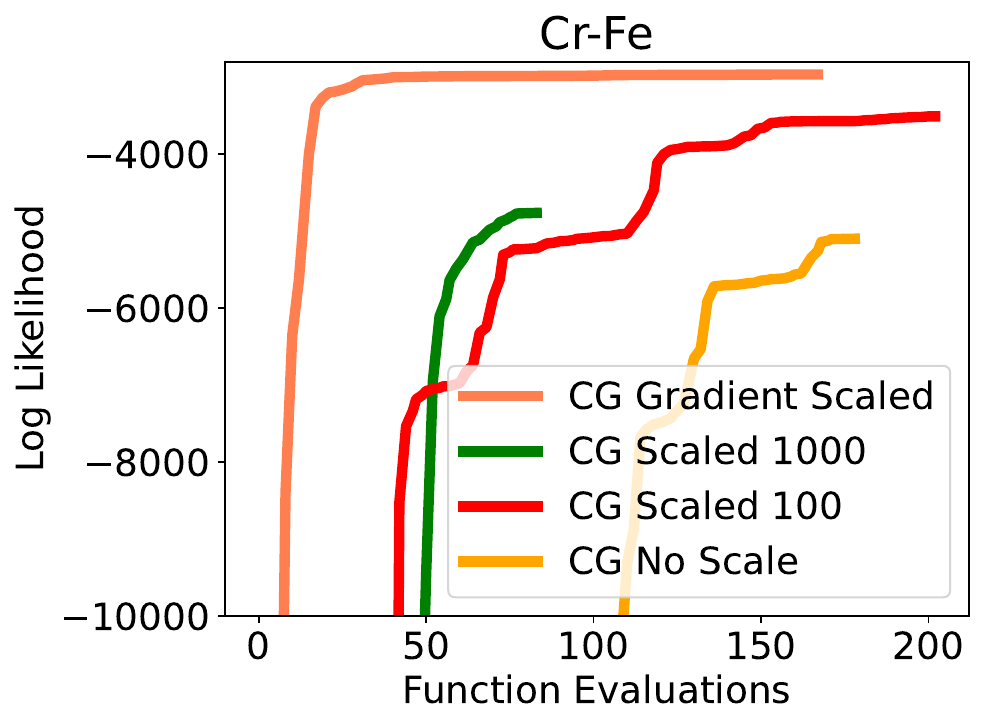}}
         \caption{Overall log-likelihood as a function of cumulative function evaluations by method for the (a-b) Cu-Mg, (c-d) Fe-Ni, (e-f) Cr-Ni, and (g-h) Cr-Fe systems. Panels in the second column provide the same information as the left for only the CG runs.}
\label{fig:efficiency_plots}
\end{figure*}

\begin{table*}[!htbp]
    \centering
    \begin{tabular}{llll}
         System & Method & Log-Likelihood & Function Evals\\
         \hline
         Cu-Mg & CG No Scale&  -4533.73& 3164\\
         &CG Scaled 100&  -4514.48& 381\\
         & CG Scaled 1000 & -4512.35 & 110\\
         &MCMC 8 CPP&  -4518.20& 42240\\
         &MCMC 4 CPP&  -4521.80& 37576\\
         &MCMC 2 CPP&  -4519.24& 43736\\
         \hline
         Fe-Ni &CG No Scale& -3632.16 & 447\\
         &CG Scaled 100& -3440.98 & 597\\
         &CG Scaled 1000& -3396.09 & 766\\
         &MCMC 8 CPP&  -3396.66 & 27720\\
         &MCMC 4 CPP& -3396.53  & 35676\\
         &MCMC 2 CPP&  -3416.54& 35946\\
         \hline
         Cr-Ni &CG No Scale& -3625.45 & 26\\
         &CG Scaled 100& -3047.24 & 316\\
         &CG Scaled 1000 (no c2 relax)& -2777.70 & 506\\
         &CG Scaled 1000 (with c2 relax)& -2764.41 & 729\\
         &MCMC 8 CPP&  -2763.35 & 39120\\
         &MCMC 4 CPP&  -2763.16 & 31800\\
         &MCMC 2 CPP& -2923.62 & 40000\\
         \hline
         Cr-Fe&CG No Scale& -5099.12 & 178\\
         &CG Scaled 100& -3509.92 & 202\\
         &CG Scaled 1000 & -4763.84 & 83\\
         &CG Gradient Scaled & -2966.95 & 167\\
         &MCMC 8 CPP& -3013.54  & 124384\\
         &MCMC 4 CPP& -2977.48  & 122824\\
         &MCMC 2 CPP& -3016.15 & 120900\\
    \end{tabular}
    \caption{Log-likelihoods corresponding to the most optimal set of parameters for all systems found by each method and the number of function evaluations needed to discover that parameter set.}
    \label{tab:combined_comparison}
\end{table*}

Fig. \ref{fig:efficiency_plots} graphically displays the overall log-likelihood as a function of cumulative objective function evaluations across all four systems and all the optimization methods used. For each MCMC iteration, the number of objective function queries is equal to the number of chains, and the recorded log-likelihood corresponds to the most optimal parameter set discovered at that iteration. Table \ref{tab:combined_comparison} provides important summary statistics from these plots, including the maximum log-likelihood attained by each method and the number of function evaluations needed to discover the corresponding optimal parameter set. 

 We will start our discussion with an examination of the MCMC runs. For each system, MCMC was employed with two, four, and eight chains per parameter (CPP). The number of iterations varied by system and number of chains, but all MCMC runs corresponding to a given system were assigned the same number of overall function evaluations. The total number of allocated function evaluations was 44,000 for Cu-Mg, 36,000 for Fe-Ni, 40,000 for Cr-Ni, and 124,800 for Cr-Fe. For Cu-Mg, we followed the guidance of the ESPEI user tutorial and used normal priors with mean and variance equal to that of the starting point value. For the other three systems, we used zero priors. 

As Fig. \ref{fig:efficiency_plots} and Table \ref{tab:combined_comparison} show, in general the runs with two chains per parameter were able to complete more iterations for a given number of function evaluations, leading to an initial performance advantage. However, as each run neared convergence, the exploratory power of more chains started nullifying the previous advantage provided by more iterations. For the Cu-Mg and Cr-Fe systems, the number of chains ended up having little influence on the optimality of the final solution or the efficiency at which that solution was determined. Conversely, the number of chains did significantly affect efficiency for Fe-Ni (eight chains per parameter reached a solution of either superior or identical optimality with approximately 8000 fewer function evaluations than the two or four chain variants). The opposite phenomenon was observed with Cr-Ni where four chains per parameter achieved an optimal solution with 7320 fewer function evaluations while the two chain variant did not appear to even converge within the allotted number of computations. When we ignore the number of computations at which the most optimal solution was found and focus on the highest log-likelihood achieved by each method within its allotted function evaluations, we see that the four or eight chain variants consistently produce most likely solutions with differences of $1.25\%$ or less. The Cu-Mg system is the only example in which the two chain variant did not produce the least optimal solution.

CG was initially performed with no parameter scaling for all systems. As Fig. \ref{fig:efficiency_plots} and Table \ref{tab:combined_comparison} illustrate, this approach generally led to premature termination of the routine due to line search failures, which were likely caused by numerical ill-conditioning. Noticing that the larger, temperature-independent coefficients barely moved throughout these processes and that their corresponding gradient component magnitudes were relatively low, we simply scaled these degrees of freedom by 100 in our second attempt. This simple preconditioning strategy resulted in significant improvements in efficiency for the Cu-Mg system and in optimality for the other three. Scaling by 1000 had a similar effect for Cu-Mg, Fe-Ni, and Cr-Ni, but resulted in a significantly less optimal solution for the Cr-Fe system. Recognizing that the current preconditioning approach was not working for the Cr-Fe system, we decided to scale all parameters such that the magnitudes of their gradients would all be on the order of 10 at the starting point. This ``Gradient Scaled" run produced a significantly more optimal result than the previous three attempts and did so with great efficiency. 

It appears as though the unscaled Cu-Mg problem is the most numerically well-conditioned of the four systems; thus, appropriate preconditioning led to efficiency advantages in finding a slightly more optimal, but similar solution. In contrast, the poor numerical conditioning of the unscaled problems for the other three systems did not allow the routine to get close to an optimal solution before terminating, leading to the apparent decrease in efficiency. But, based on the initial slopes and the values reached before slowing by the efficiency curves plotted in Fig. \ref{fig:efficiency_plots}, we can speculate that the preconditioned runs that discovered more optimal solutions did so in a more efficient manner than their unscaled counterparts would have based on their steeper initial slopes.

Examination of the line search queries at the end of the Scaled 1000 run for the Cr-Ni system showed that solutions with larger log-likelihoods were being found, but the algorithm was not stepping. Relaxing the $c2$ curvature condition from $c2=0.4$ to $c2=0.6$ led to a slight increase in optimality, but at a large cost in function evaluations. We believe this is due to a plateau-like region near the optimum in the log-likelihood surface. That is, the algorithm could not find a step size which produced sufficient flattening of the directional derivative because the slope was already quite flat, and small derivative magnitudes led to small steps even with reasonably large step sizes. 

\begin{figure*}[!htbp]
        \captionsetup[subfigure]{labelformat=empty}
        \centering
        \subfloat[(a)]{\includegraphics[width=.5\linewidth]{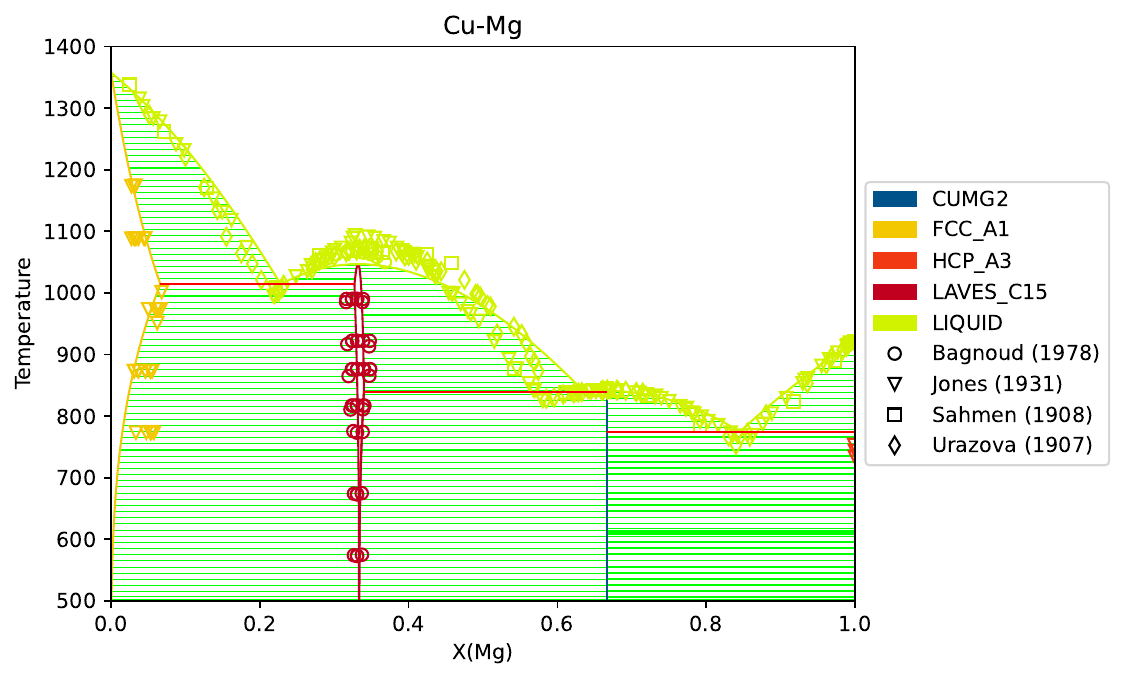}}
         \subfloat[(b)]{\includegraphics[width=.5\linewidth]{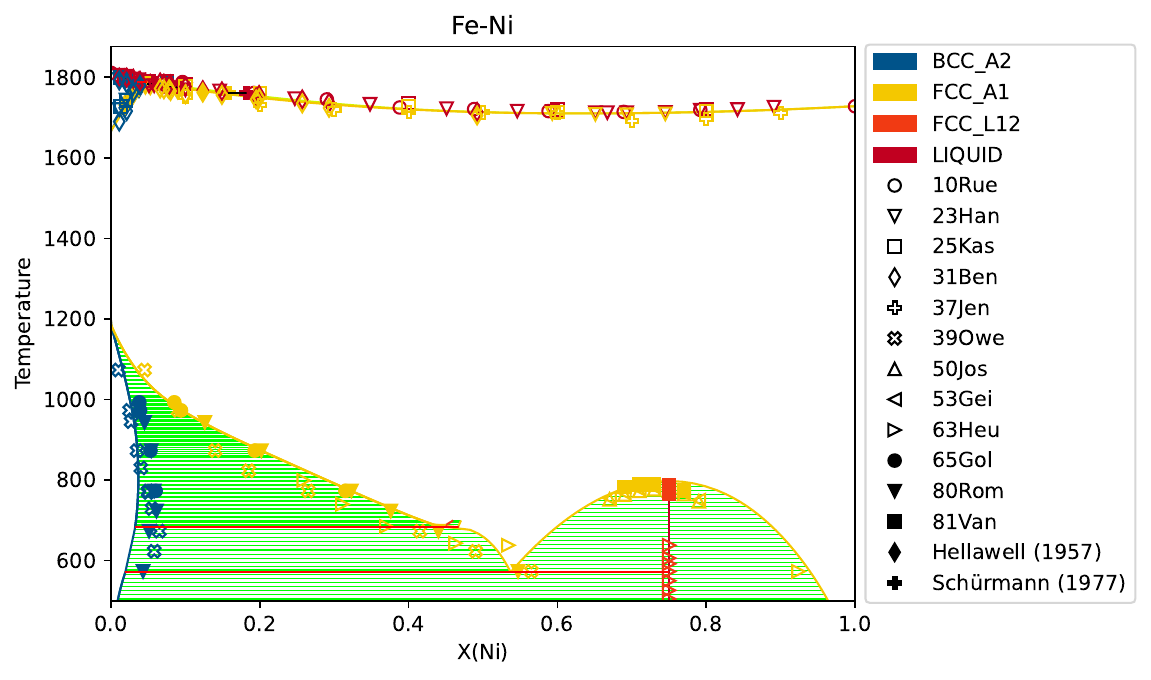}}\\
         \subfloat[(c)]{\includegraphics[width=.5\linewidth]{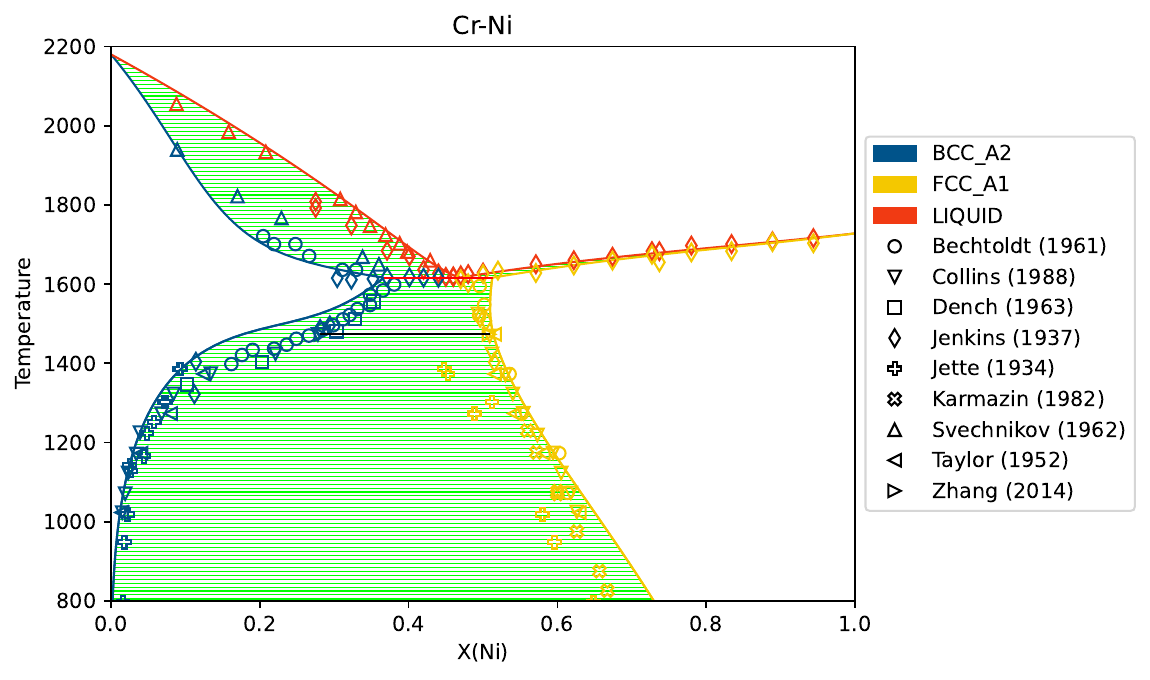}}
         \subfloat[(d)]{\includegraphics[width=.5\linewidth]{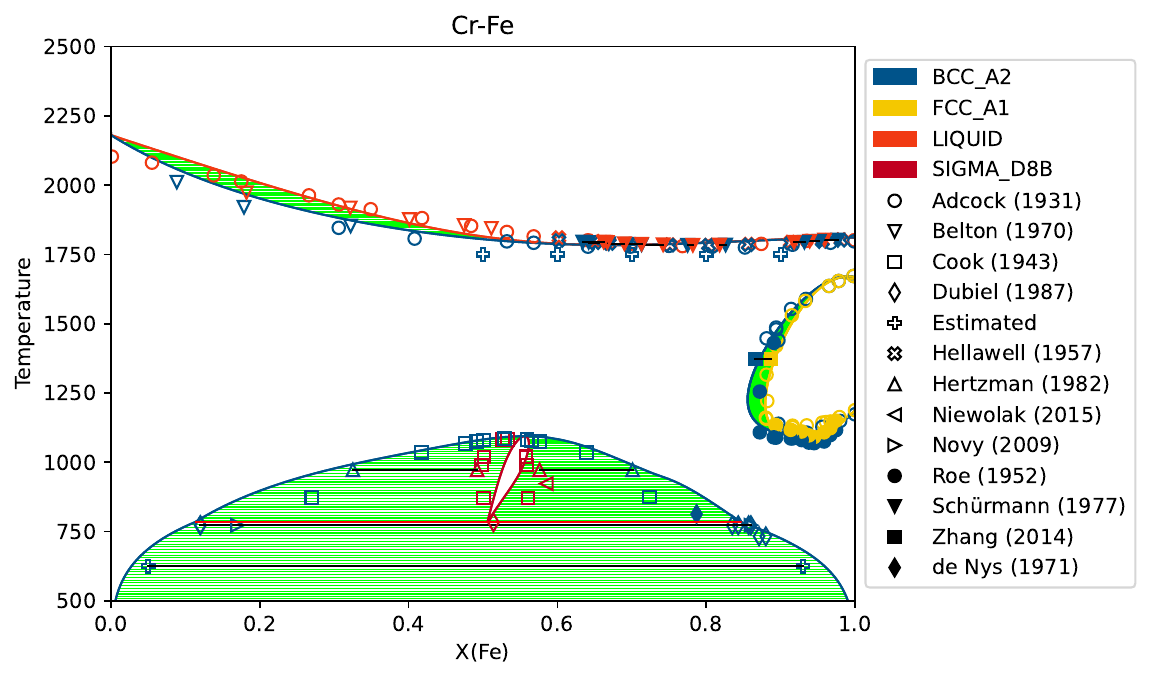}}\\
         \caption{Phase diagrams of the (a) Cu-Mg, (b) Fe-Ni, (c) Cr-Ni, and (d) Cr-Fe systems corresponding to the most optimal solution found through CG.}
\label{fig:phase diagrams}
\end{figure*}

Phase diagrams produced from the most optimal solution found using CG for all four systems are displayed in Fig. \ref{fig:phase diagrams}. The Cu-Mg, Fe-Ni, and Cr-Ni models show excellent agreement with all experimental phase equilibria data, and the Fe-Ni diagram exhibits the Nishizawa horn feature. We also see fantastic agreement in the Cr-Fe phase diagram, but we note that the solubility of the sigma phase is rather narrow in comparison to the data. Using the same sublattice model as the current authors, Jacob et al. \cite{jacob2018} recently achieved wider sigma phase solubility partly through the addition of two sigma phase interaction parameters. While it is beyond the scope of this work to argue new assessments for these systems, improvement in the Cr-Fe sigma phase solubility could be made by including first-principles calculations of sigma phase mixing enthalpies in the parameter generation and fitting steps. We also acknowledge that, as models get more complicated, more nonlinearities are likely to be introduced into the likelihood function, which can result in multiple local maxima. It is possible that there exists a more likely parameter set which addresses the sigma phase solubility given our current model, but that our local optimizer was pulled into a local maximum based on the starting point.

Starting the method comparison with computational efficiency, it is evident from Fig. \ref{fig:efficiency_plots} and Table \ref{tab:combined_comparison} that all CG runs (even those which terminated early due to numerical ill-conditioning), significantly out-performed all MCMC variants. Indeed, as a conservative estimate of improvement in computational efficiency, if we add all of the function evaluations needed to test all preconditioning strategies for each system and compare that to the number of function evaluations computed by the more efficient of the four or eight chain MCMC variant, we see that CG was approximately 10, 15, 30, and 195 times more efficient for Cu-Mg, Fe-Ni, Cr-Ni, and Cr-Fe, respectively. This implies that, even if an appropriate preconditioning strategy is not known ahead of time, the power of gradient-based methods can still allow for a significant improvement in computational efficiency. Or, multiple starting points could be attempted if there is concern about falling into local maxima. For the case of an expert with \textit{a priori} knowledge of an effective preconditioning approach, we can consider only the number of function evaluations pertaining to the CG run which attained the most optimal solution. If we consider the same number of function evaluations for MCMC, this scenario provides computational efficiency multipliers for CG over MCMC of approximately 341, 36, 44, and 735. Furthering this scenario, examination of Table \ref{tab:combined_comparison} shows that, for the Cu-Mg and Cr-Fe systems, no MCMC variant quite reached the optimality of the CG Scaled 1000 and CG Gradient Scaled runs, respectively. The number of function evaluations necessary for these CG runs to reach the most optimal likelihoods achieved by the more efficient MCMC variant (-4521.80 and -2977.48) were 40 and 105, resulting in computational efficiency multipliers of 939 and 1170, respectively. Of course, because each MCMC run consumes all computational resources necessary to perform all of the allocated iterations/function evaluations, these multipliers are still somewhat conservative. We note that the multipliers for the higher-dimensional problems, those of Cu-Mg and Cr-Fe, are significantly higher than those of Fe-Ni and Cr-Ni for the case of an \textit{a priori} preconditioning strategy, leading us to speculate that even greater efficiency gains could be realized when fitting more complicated binary or even ternary systems.

Further inspection of Fig. \ref{fig:efficiency_plots} and Table \ref{tab:combined_comparison} reveals that CG did not trade optimality for these large gains in efficiency. Indeed, if we compare the most optimal solutions found by both methods for each system, we see differences in log-likelihood of less than one percent. This shows that gradient-based methods have the potential to greatly increase the computational efficiency of the CALPHAD model parameter calibration process without sacrificing accuracy of the final solution.

\section{Conclusions and Future Work}

In this work, we implemented and demonstrated an analytic gradient-based framework for the optimization of CALPHAD model parameters. This framework is flexible enough to be applied to arbitrarily-complex models and residual functions which require any number of energy minimization calculations. It can also be used with any off-the-shelf or custom-built gradient-based optimization method. The key to our approach is the adaptation of the Jansson derivative technique to treat model parameters as conditions of the equilibrium calculation.

Our demonstration examined the use of CG to optimize the parameters for four binary systems of varying complexity. We showed that with simple preconditioning strategies, all four systems could be optimized with only hundreds of objective function evaluations. This is an enormous improvement in computational efficiency over black-box methods such as MCMC which was demonstrated to require tens to hundreds of thousands of function evaluations to achieve results of competitive optimality. We do admit that, while it was simple and effective, our preconditioning approach is likely less than optimal. Looking forward, the capability to compute second-order Jansson derivatives would provide a computationally efficient way of calculating analytic Hessians (i.e. requiring no further objective function evaluations or matrix inversions), and this information could be dynamically injected into the optimizer to re-condition following each iteration. Second-order Jansson derivatives would also open the door to gradient-based techniques such as Newton's method, which leverage Hessian information for improved performance. Next steps could also include considering global optimization wrappers for local gradient-based techniques such as basin hopping or branch and bound. However, these methods still do not address the uncertainty quantification and propagation problem, which we believe is a crucial aspect of properly understanding and employing CALPHAD models. Hamiltonian Monte Carlo, a gradient-informed, global, stochastic optimization method, would be a natural extension of ESPEI's current MCMC standard. We anticipate that this gradient-based method would be able to provide improvements not only in computational efficiency, but also in solution robustness, while also facilitating uncertainty quantification.



\section*{Acknowledgments}

Work at Lawrence Livermore National Laboratory (LLNL) was performed under the auspices of the U.S. Department of Energy by LLNL under Contract No. DE-AC52-07NA27344. RA also wishes to acknowledge the support from NSF through Grant No. NSF-DMREF-2119103.

\bigskip
\bibliography{mybib} 

    \appendix 
    \counterwithin*{figure}{section}
    \counterwithin*{table}{section}
    
    \section{}\label{sec:appendix}

    \begin{table}[!htbp]
    \begin{minipage}{\textwidth}
    \centering
    \begin{tabular}{llccc}
       Phase  & Parameter & Starting Point & CG Best & MCMC Best\\
       \hline
       Liquid & $^0L^{Liquid}_{Cu, Mg}$  & $-34437.6 + 11.428T$ &$-34219.1+11.698T$ &$-34301.8+11.717T$\\
         & $^1L^{Liquid}_{Cu, Mg}$  & $-8720.5 + 5.106T$ & $-9500.5+4.529T$&$-8916.8+3.240T$\\
       FCC  & $^0L^{FCC}_{Cu, Mg}$ & $-14675.0 + 4.892T$ & $-14886.0+4.912T$ &$-14732.1+5.178T$\\
         & $^1L^{FCC}_{Cu, Mg}$ & $8236.3 - 2.744T$ & $7138.0-2.511T$ & $7736.3-3.330T$\\
       HCP  & $^0L^{HCP}_{Cu, Mg}$ & $20149.6$ & $20149.6$ &$18873.5$\\
         & $^1L^{HCP}_{Cu, Mg}$ & $-24441.2$ & $-24441.0$&$-23277.2$\\
       CuMg\textsubscript{2}  & $^0G^{CUMG2}_{Cu:Mg}- (^0G^{FCC}_{Cu} + 2{}^0G^{HCP}_{Mg})$ & $-32539.5 + 13.2T$ & $-32866.6+13.052T$ & $32888.3+12.583T$\\
       Laves C15 & $^0G^{Laves}_{Cu:Cu}-3{}^0G^{FCC}_{Cu}$& $46500.0 -15.501T$ & $45152.7-15.656T$ &$44547.3-15.048T$\\
        & $^0G^{Laves}_{Mg:Mg}-3{}^0G^{HCP}_{Mg}$ & $21000.0 -6.999T$ & $21938.3-6.896T$&$21900.5-6.711T$\\
         & $^0G^{Laves}_{Cu:Mg}-(2{}^0G^{FCC}_{Cu} + {}^0G^{HCP}_{Mg})$ & $-39591.3 + 15.72T$ & $-42195.3+14.154T$ & $-42948.0+14.390T$\\
          & $^0G^{Laves}_{Mg:Cu}-(^0G^{FCC}_{Cu} + 2{}^0G^{HCP}_{Mg})$ & $104160.0 -34.719T$ &$104159.6-34.719T$ &$104448.9-34.345T$\\
          & $^0L^{Laves}_{Cu:Cu,Mg}$ & $17772.0$ &$2391.9$ &$7292.8$\\
          & $^0L^{Laves}_{Cu,Mg:Mg}$ & $21240.0$ &$35694.7$ &$35668.0$\\
          
    \end{tabular}
    \caption{Generated model and starting point as well as best parameter fits from CG and MCMC methods for the Cu-Mg system.}
    \label{tab:cu_mg_details}
    \end{minipage}
\end{table}

\begin{table}[!htbp]
\begin{minipage}{\textwidth}
    \centering
    \begin{tabular}{llccc}
       Phase  & Parameter & Starting Point & CG Best & MCMC Best\\
       \hline
Liquid & $^0L^{Liquid}_{Fe, Ni}$  & $-17866.8$&$-17694.9$ &$-17527.2$\\
       & $^1L^{Liquid}_{Fe, Ni}$  & $12618.6$& $11330.7$&$11370.3$\\
BCC & $^0L^{BCC}_{Fe, Ni}$  & $-8584.6+9.585T$& $590.8-3.811T$&$562.5-3.638T$\\
FCC & $^0L^{FCC}_{Fe, Ni}$  & $-13536.9+1.168T$&$-12249.6-2.156T$ &$-12336.1-2.026T$\\
    & $^1L^{FCC}_{Fe, Ni}$  & $12375.2$& $11463.6$&$11548.2$\\
FCC L1\textsubscript{2} & $^0G^{L1_2}_{Fe:Ni}-({}^0G^{BCC}_{Fe} + 3{}^0G^{FCC}_{Ni})$ & $-31398.1+3.400T$ &$-61118.5+36.493T$ & $-61269.1+36.634T$
       \end{tabular}
    \caption{Generated model and starting point as well as best parameter fits from CG and MCMC methods for the Fe-Ni system.}
    \label{tab:fe_ni_details}
    \end{minipage}
\end{table}

\clearpage

\begin{table}[!htbp]
\begin{minipage}{\textwidth}
    \centering
    \begin{tabular}{llccc}
       Phase  & Parameter & Starting Point & CG Best & MCMC Best\\
       \hline
Liquid & $^0L^{Liquid}_{Cr, Ni}$  & $-13886.1+4.632T$& $-14080.6+9.363T$&$-14190.5+9.339T$\\
BCC & $^0L^{BCC}_{Cr, Ni}$  & $9545.6-4.222T$& $9436.9+0.322T$ &$11118.2-0.807T$\\
& $^1L^{BCC}_{Cr, Ni}$  & $47004.2-11.221T$ & $33296.0-13.054T$&$31119.9-11.689T$\\

FCC & $^0L^{FCC}_{Cr, Ni}$  & $10224.7-8.477T$ & $3576.5-2.210T$&$3465.5-2.231T$\\
    & $^1L^{FCC}_{Cr, Ni}$  & $36724.5-15.324T$ & $14673.1-9.526T$&$15565.7-10.051T$\\
       \end{tabular}
    \caption{Generated model and starting point as well as best parameter fits from CG and MCMC methods for the Cr-Ni system.}
    \label{tab:cr_ni_details}
    \end{minipage}
\end{table}

\begin{table}[!htbp]
\begin{minipage}{\textwidth}
    \centering
    \begin{tabular}{llccc}
       Phase  & Parameter & Starting Point & CG Best & MCMC Best\\
       \hline
Liquid & $^0L^{Liquid}_{Cr, Fe}$  & $-1977.98-2.0T$ & $-2130.9-2.0T$&$-3318.5+0.3T$\\
BCC & $^0L^{BCC}_{Cr, Fe}$  & $25460.6-12.0T$ & $23840.5-12.1T$ & $20434.9-8.6T$\\
& $^1L^{BCC}_{Cr, Fe}$  & $-5953.99-1.5T$ & $-2137.6+3.1T$& $-1255.0+2.4T$\\
FCC & $^0L^{FCC}_{Cr, Fe}$  & $10000.0-8.0T$ & $12114.5-6.8T$&$10628.0-6.7T$\\
    & $^1L^{FCC}_{Cr, Fe}$  & $-14248.9-20.7T$ & $-1501.1+5.7T$& $2587.5+0.03T$\\
$\sigma$ & ${}^0G^{\sigma}_{Fe:Fe:Fe}-3{}^0G^{BCC}_{Fe}$ & $427769.0-77.4T$ & $316905.2-170.6T$ & $296366.4-112.8T$\\
& ${}^0G^{\sigma}_{Fe:Fe:Cr}-(2{}^0G^{BCC}_{Fe}+{}^0G^{BCC}_{Cr})$ & $234492.0-82.2T$ & $59787.5-127.5T$ & $58951.1-132.2T$\\
& ${}^0G^{\sigma}_{Fe:Cr:Fe}-(2{}^0G^{BCC}_{Fe}+{}^0G^{BCC}_{Cr})$ & $373713.0-75.0T$ & $328235.3-125.0T$ &$196577.9+59.1T$\\
& ${}^0G^{\sigma}_{Cr:Fe:Fe}-(2{}^0G^{BCC}_{Fe}+{}^0G^{BCC}_{Cr})$ & $407175.0-108.9T$ & $401069.8-114.6T$ & $745881.0-164.2T$\\
& ${}^0G^{\sigma}_{Fe:Cr:Cr}-({}^0G^{BCC}_{Fe}+2{}^0G^{BCC}_{Cr})$ & $194605.0-54.9T$ & $193440.3-55.9T$ & $201777.1-146.4T$\\
& ${}^0G^{\sigma}_{Cr:Fe:Cr}-({}^0G^{BCC}_{Fe}+2{}^0G^{BCC}_{Cr})$ & $481533.0-177.9T$ & $471403.0-184.0T$ &$579980.6-83.1T$\\
& ${}^0G^{\sigma}_{Cr:Cr:Fe}-({}^0G^{BCC}_{Fe}+2{}^0G^{BCC}_{Cr})$ & $409588.0-126.0T$ & $397823.4-138.8T$ & $610113.8-279.5T$\\
& ${}^0G^{\sigma}_{Cr:Cr:Cr}-3{}^0G^{BCC}_{Cr}$ & $398349.0-137.2T$ & $375473.2-165.7T$ & $544415.4-240.5T$\\
       \end{tabular}
    \caption{Generated model and starting point as well as best parameter fits from CG and MCMC methods for the Cr-Fe system.}
    \label{tab:cr_fe_details}
    \end{minipage}
\end{table}

\end{document}